\documentclass[sigconf]{acmart}

\renewcommand\footnotetextcopyrightpermission[1]{} 
\usepackage{algorithm}
\usepackage[noend]{algpseudocode}
\usepackage{booktabs} % For formal tables
\usepackage{float}
\usepackage{graphicx}
\usepackage{caption}
\usepackage{longtable}
\usepackage{ltablex}
\usepackage{multicol}
\usepackage{subcaption}
\usepackage{supertabular}
\usepackage{tabularx} % in the preamble
\usepackage{textcomp}
\usepackage{url}

\setcopyright{rightsretained}
\begin{document}
\title{University Twitter Engagement: Using Twitter Followers to Rank Universities}

\author{Corren G. McCoy}
\orcid{}
\affiliation{%
	\institution{Old Dominion University}
	\streetaddress{Computer Science Department}
	\city{Norfolk} 
	\state{VA} 
	\postcode{23529}
}
\email{cmccoy@odu.edu}

\author{Michael L. Nelson}
\affiliation{%
	\institution{Old Dominion University}
	\streetaddress{Computer Science Department}
	\city{Norfolk} 
	\state{VA} 
	\postcode{23529}
}
\email{mln@cs.odu.edu}

\author{Michele C. Weigle}
\affiliation{%
	\institution{Old Dominion University}
	\streetaddress{Computer Science Department}
	\city{Norfolk} 
	\state{VA} 
	\postcode{23529}
}
\email{mweigle@cs.odu.edu}

\renewcommand{\shortauthors}{C. McCoy et al.}

\begin{abstract}
We examine and rank a set of 264 U.S. universities extracted from the National Collegiate Athletic Association (NCAA) Division I membership and global lists published in U.S. News, Times Higher Education, Academic Ranking of World Universities, and Money Magazine. Our University Twitter Engagement (UTE) rank is based on the friend and extended follower network of primary and affiliated secondary Twitter accounts referenced on a university\textquotesingle s home page. In rank-to-rank comparisons we observed a significant, positive rank correlation ($\tau$=0.6018) between UTE and an aggregate reputation ranking which indicates that UTE could be a viable proxy for ranking atypical institutions normally excluded from traditional lists. In addition, we significantly reduce the cost of data collection needed to rank each institution by using only web-based artifacts and a publicly accessible Twitter application programming interface (API).
\end{abstract}

\maketitle

\section{Introduction}

Universities and other academic institutions increasingly see their presence, visibility and footprint on the Web as central to their reputation and international standing. In this context, the academic web is evolving into more than a vehicle for communicating scientific and cultural achievements; information content is viewed as a reflection of the overall organization and performance of the university~\cite{aguillo2008webometric}. Academic rankings, therefore, play an important role in assessing reputation. With different criteria and disparate methodologies, there can be a significant divergence in the rankings of a particular institution depending upon the list that is surveyed.

Academic excellence is difficult to quantify, yet most ranking organizations start by collecting performance indicators (e.g., Nobel laureates, research volume) about each university which they believe to be independent indicators of quality. After giving each a different, predetermined weight, the indicators are summed to a total score that determines the  university\textquotesingle s rank. The weighted scoring method is sometimes supplemented with a peer institution survey which is compiled and submitted by academic experts ~\cite{enserink2007ranks}. We propose an alternative metric for ranking universities, University Twitter Engagement (UTE), a score which is the sum of all affiliated users the university promotes on its homepage plus the followers of any Twitter friends who indicate an affiliation with the university in their profile Uniform Resource Identifier (URI). The UTE score is an important metric as it quantifies the potential popularity or prestige of the university without an extensive data collection effort.

\begin{figure} [h]
	\begin{subfigure}[b]{0.5\textwidth}
		\includegraphics[width=\textwidth, height=\textheight, keepaspectratio]{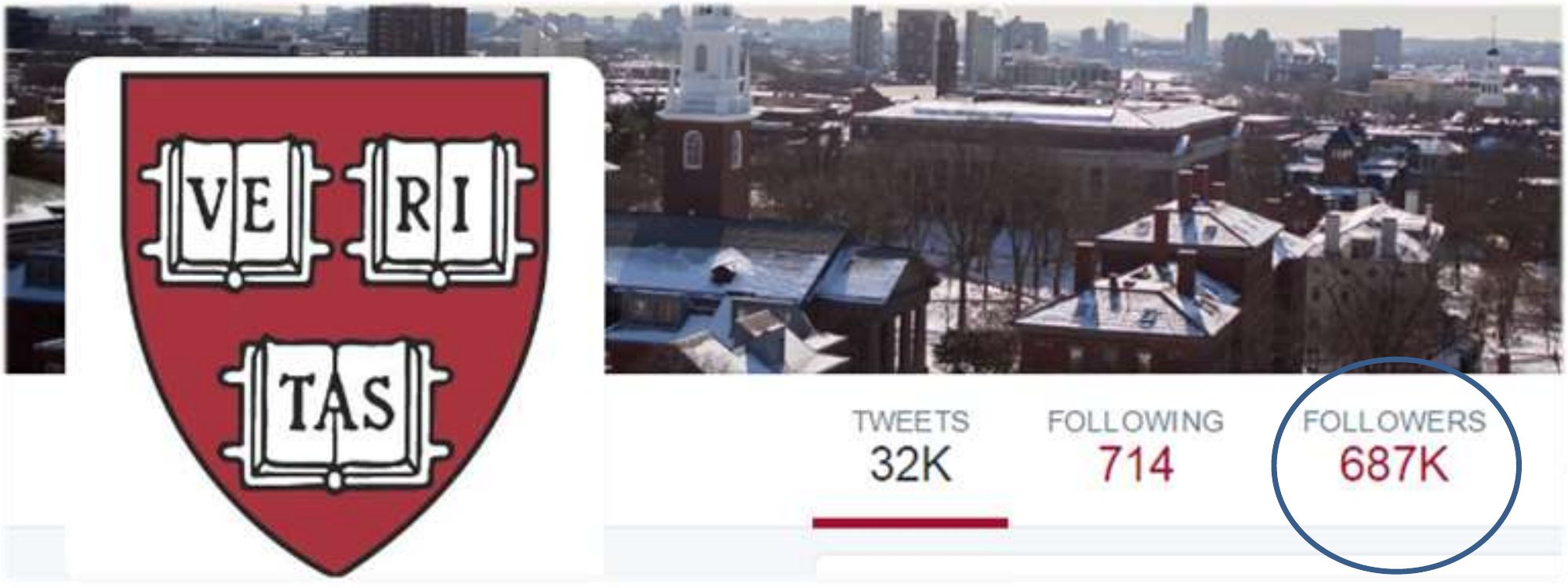}
		\caption{Harvard University @HARVARD.}
		\label{fig:Twitter Profile for @Harvard.}
	\end{subfigure}
	\begin{subfigure}[b]{0.5\textwidth}
		\includegraphics[width=\textwidth, height=\textheight, keepaspectratio]{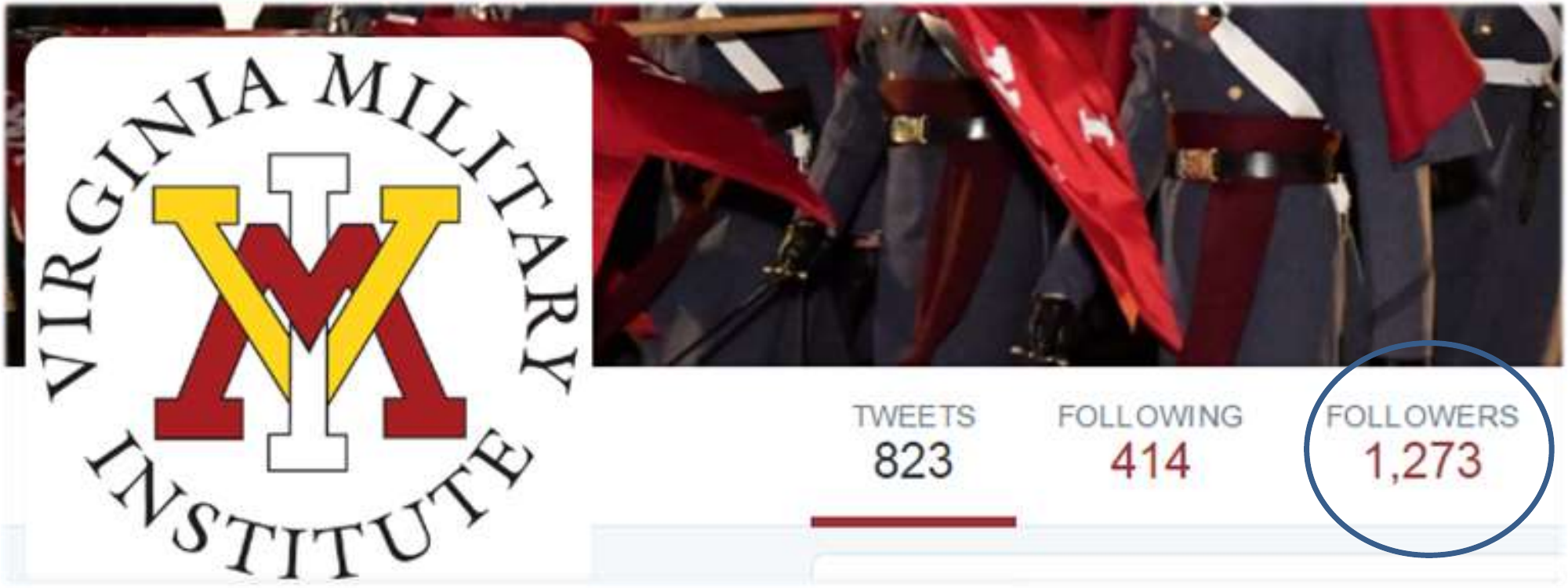}
		\caption{Virginia Military Institute @VMILife.}
		\label{fig:Twitter Profile for @VMILife}
	\end{subfigure}
	\caption{Twitter Follower Comparison}\label{fig:TwitterFollowerComparison}
\end{figure}

This research assumes that (1) universities with higher undergraduate enrollment are likely to have more Twitter followers as students graduate and transition to alumni status, (2) official Twitter accounts will be well advertised on the university\textquotesingle s homepage, (3) sports participation is a driver that increases awareness of the university\textquotesingle s brand, and (4) the data needed to comprise the ranking criteria is readily available and easy to collect from public data sources on the web. Figure ~\ref{fig:TwitterFollowerComparison} depicts a recent glimpse into the Twitter followers (675K) for Harvard University, a perennially top-ranked school, which represents an approximate 100:1 ratio to its undergraduate enrollment (6,660). On the other hand, the Twitter follower count (1,213) for Virginia Military Institute (VMI), a top 100 school, barely maintains a 1:1 ratio with its undergraduate enrollment (1,717). If we only consider alumni, we would expect that schools with similar enrollment would attract a similar number of Twitter followers. The large disparity between Harvard and VMI presents a first indication that some correlation may exist between rank position and Twitter followers. We propose a novel approach which considers not only the primary Twitter accounts which the university may advertise on its homepage, but secondary accounts which the university informally promotes by following them on Twitter. In order to ensure that a relationship or mutual affiliation exists between the primary and secondary accounts, we enforce the requirement that the top level domain assigned to the university in its URI (e.g., harvard.edu) must be present in the Twitter profile of all affiliated Twitter accounts.
                                      
The contributions of this study are as follows:
\begin{itemize}
	\item We aggregate the rankings from multiple expert sources to calculate an adjusted reputation rank (ARR) for each university which allows direct comparison based on position in the list and provides a collective perspective of the individual rankings.
	\item We conduct a web-based analysis to identify and collect a mutually aligned, comprehensive set of primary and secondary Twitter accounts as a measure of social media engagement.
	\item We propose an easily collected proxy measurement, UTE, that achieves comparable rankings as more complex methodologies which rely upon manual compilation.
	\item We produce a social media rich dataset  containing Twitter profile data and  institutional demographics which will reduce the effort required by other researchers to reproduce our work ~\cite{weller2016manifesto}. The complete dataset is posted on GitHub\footnote{\url{https://github.com/oduwsdl/University-Twitter-Engagement}}.
\end{itemize}

\section{Related Work}
The relevance of Twitter followers as a means of measuring reputation has been the subject of many previous studies. Our work parallels the studies conducted by Klein et al. ~\cite{klein2009comparing, klein2009correlation} and Nelson et al. ~\cite{correlation:techreport:2008} who attempt to find correlations between the rankings of real-world entities (e.g., college football teams, Billboard Hot 100, graduate business schools) and the page rank of their respective home pages. In this paper, we examine something similar, but instead derive the ranking score using social media.

\subsection{The Challenge of Ranking Universities}
University rankings are subject to normative assumptions about the type of variables used and their associated weightings. Therefore, ranking systems reflect the conceptual framework and the modeling choices used to build them ~\cite{goglio2016one}. These systems can potentially give inaccurate indications to university administrators about the activities in which it is better to invest in order to improve the ranking of their institution ~\cite{goglio2016one}. And, as predicted by decision-making theory, Bowman and Bastelo \cite{bowman2011anchoring} found that anchoring effects exert a substantial influence on future reputational assessments. Once a university reaches the pinnacle of any ranking system, they are \textit{anchored} and often do not fall very far from their original position. Bowman and Bastelo ~\cite{bowman2011anchoring} observed that academics across the world are influenced in some way by external assessments of their ranking. Further, they concluded it would take an extensive change in academic quality to significantly influence reputation scores in any given year. Nearly always, rankings drive reputation, not the other way around. The notion of reputation largely serves as a feedback loop to maintain the status quo, establishing the credibility of the rankings and ensuring stability in results over time ~\cite{bowman2011anchoring}.

Different metrics used by the ranking organizations can make direct comparisons difficult as each list may be intended to convey a distinct purpose. Three of the four ranking systems referenced in this paper determine best colleges based on academic excellence while the fourth, Money Magazine, is focused solely on perceived value and affordability. A particular ranking list may count factors such as external funding, numbers of articles and books authored by faculty members, library resources, proportion of faculty members with advanced degrees, and quality of students based on admissions criteria. With so many heterogeneous metrics, conducting surveys can be time consuming and expensive if the data must be gathered over a long period of time or requires manual input from a university official. These numbers are not easy to obtain and are assumed to be an adequate proxy for quality.   

The assumption by the ranking systems is that one set of metrics can be applied to every institution and that the norms of research-based and elite universities are the gold standard that can be applied to everyone ~\cite{altbach2015dilemmas}. Goglio ~\cite{goglio2016one} showed that the competition to improve ranks among lower ranked universities is different from the competition to do so among higher ranked universities. The rank-localized nature of competition is primarily among those universities that are similarly ranked. Grewal et al.\textquotesingle s ~\cite{grewal2012university} results also showed that a top-ranked university has a 0.965 probability of finishing in the top five the next year.  Ultimately, regardless of popularity, universities exhibit very little power to control their rank position and, although almost all aspire to be among the upper echelons, the top positions are perennially dominated by the same institutions ~\cite{goglio2016one}.

\subsection{Social Media in Higher Education}
Even when the ranking systems have the same goal, technical challenges can still hamper data collection; specifically, changes in page names or web domains can affect both the visibility and discoverability of the institution\textquotesingle s web presence. An organization can also use different web domains for search engines, aliases and independent domains for some of their subunits or services ~\cite{aguillo2008webometric}. For example, in addition to \textit{odu.edu} which is the expected domain for Old Dominion University, we found \textit{odu.trisigma.org} and \textit{oduwsoccerclub.wixsite.com} as domains associated with university-sponsored clubs. As noted by Aguillo ~\cite{aguillo2008webometric}, an adequate web presence or lack thereof may not always correlate with the quality or prestige of the institution.

Social networking sites have  proven to be an effective vehicle for organizations seeking to implement diverse branding strategies, given that such sites allow consumers to share their experiences and opinions concerning the organization\textquotesingle s products and brand in real time ~\cite{heller2011social, jansen2009twitter}. Many organizations have rapidly adopted social networking services such as Facebook and Twitter, a move that has altered the face of customer relationship management from managing customers to collaborating with customers. While social media interactions in the higher education space are not transactional in the traditional sense, they do provide a way for institutions to continually engage with their constituents. Another form of engagement, or public involvement with a chosen organization that may fall outside of consumer interests is \emph{affective commitment} which Kang ~\cite{kang2014understanding} defines as a voluntary bonding between entities; perhaps similar to how a university might maintain contact with its alumni long after graduation. We will focus on engagement at a very basic or minimal level based on familiarity and cognition where one first needs to be familiar with a university\textquotesingle s online activity and subsequently start to follow them via social media.

As part of their ongoing research to measure the impact and social media usage in the United States, a 2016 study conducted by the Pew Research Center concluded that while Facebook continues to be the U.S.\textquotesingle s most popular social networking site with nearly 79\% of online users using the platform, Twitter usage is holding steady at 24\% and is also somewhat more popular among the highly educated ~\cite{greenwood2016social}. Go et al.\textquotesingle s ~\cite{go2016but} 2016 social media benchmarking report also suggests that Twitter is perceived as the most useful application for businesses.  At the organizational level Tsimonis et al. ~\cite{tsimonis2014brand} examined the policies, strategies and outcomes that companies might expect when engaging on social media. One observed outcome related to increased brand awareness theorized that it is possible to use a well-designed webpage to spark additional interest. Further, research findings attest to the value of social media engagement in building communities and nurturing positive public attitudes regarding the reputation of the organization ~\cite{men2015infusing}. Through data collected via a large scale survey Dikjmans et al. ~\cite{dijkmans2015stage} also found that engagement in social media activities is positively related to corporate reputation.

\subsection{Influence of Twitter Followers}
Measuring  influence  and  social  networking  potential  on  Twitter has been discussed in various papers as well as in numerous blogs  and  online  media.  Related  scientific  work  on  Twitter includes approaches which measure influence by not only taking followers  and  interactions  into  account,  but  also  by  analyzing topical  similarities  with  the  help of a ranking method  similar to PageRank ~\cite{weng2010twitterrank}. Other approaches define different types of influence on Twitter, namely indegree, retweet and mention influence ~\cite{cha2010measuring}. Accordingly, a question that arises concerns how to determine the Twitter accounts that are most influential and how their influence is subsequently measured ~\cite{antoniadis2016method}. Measuring Twitter followers is generally considered to be a popular metric as having many followers can indicate a higher level of influence as more people seem to be interested in the user. This metric implies that the more followers a user has, the more impact the user has, as the user seems to be more popular ~\cite{leavitt2009influentials}. Preussler ~\cite{preussler2010managing} contends that the number of followers is an indicator for the social reputation and the number of followers will increase as the user becomes more important. Finally, Kunegis et al. ~\cite{Kunegis:2013:PAO:2464464.2464514} assert that preferential attachment indicates that people who already have many ties are more likely to receive new ties. In other words, people who are followed by many people (i.e., are popular) are more likely to receive new followers.

An alternative approach for ranking Twitter users undertaken by Saito and Masuda ~\cite{saito2013two} considers the number of others that a user follows, i.e. friends. They concluded that the number of others that a user follows is equally important as the number of followers when estimating the importance of a Twitter user. In previous studies on Twitter, a variety of characteristics, both personal and social, have been used to identify influencers and each study measures influence from different perspectives ~\cite{leavitt2009influentials,black1993evaluating,kwak2010twitter,weng2010twitterrank}.  Weng introduced the concept of  \textit{homophily} which implies that a Twitterer follows a friend because she is interested in some topics the friend is publishing, and the friend follows back because she finds they share a similar topical interest. The presence of homophily implies there are Twitter users who are highly selective when choosing friends to follow ~\cite{weng2010twitterrank}.  These conclusions are evidenced by super users who are followed by many other users, but do not follow back equally as they only follow a select group of Twitter friends or other super  users (e.g., consider the friend-to-follower ratio of Harvard shown in Figure ~\ref{fig:TwitterFollowerComparison}).

\section{Methodology}
The following section discusses how we chose the performance indicators to correspond with the entries in the expert lists, the ranking algorithm and other operational details.

\subsection{Establishing the Selection Criteria}
To select the universities of interest, we begin with the 351 American colleges and universities currently classified as Division I by the National Collegiate Athletic Association\footnote{\url{http://www.ncaa.org/about?division=d1}} (NCAA). We then consider which of these institutions appear among the rankings of the Academic Rankings of World Universities\footnote{\url{http://www.shanghairanking.com/}} (ARWU) 2016, the Times Higher Education\footnote{\url{https://www.timeshighereducation.com/world-university-rankings/2016/world-ranking}} (THE) World University rankings 2015-2016, Money\textquotesingle s Best Colleges\footnote{\url{http://new.time.com/money/best-colleges/rankings/best-colleges/}} (MONEY) 2016-2017, and U.S. News (USNEWS) Best Global Universities\footnote{\url{http://www.usnews.com/education/best-global-universities}} 2015 and 2016. 

% \usepackage{booktabs}
% \usepackage{graphicx}
\begin{table}[ht]
	\centering
	\caption{Contribution of Each Ranking List to Our Dataset}
	\label{tab:University Ranking Lists}
	\resizebox{0.9\columnwidth}{!}{%
		\begin{tabular}{@{}lccc@{}}
			\toprule
			\multicolumn{1}{c}{\begin{tabular}[c]{@{}c@{}}Ranking \\ System\end{tabular}} & \begin{tabular}[c]{@{}c@{}}Total\\ Universities\end{tabular} & \begin{tabular}[c]{@{}c@{}}U.S. \& NCAA\\ Division 1\end{tabular} & \begin{tabular}[c]{@{}c@{}}Unique\\ Entries\end{tabular} \\ \midrule
			ARWU & 500 & 107 & 1 \\
			Money Magazine & 705 & 249 & 115 \\
			THE & 800 & 118 & 4 \\
			US News 2015 & 500 & 99 & 0 \\
			US News 2016 & 750 & 137 & 3 \\
			Any Two Lists & -- & -- & 22 \\
			Any Three Lists & -- & -- & 19 \\
			Any Four Lists & -- & -- & 16 \\
			All Five Lists & -- & -- & 84 \\\bottomrule
			&  & Total & 264 \\ 
		\end{tabular}%
	}
\end{table}

In Table ~\ref{tab:University Ranking Lists}, we identify the overlap between the total number of universities on each list and the NCAA Division I category of interest to our study. While Division I is not necessarily a ranking, participation in Division I is an indicator that the university has a vested interest in engaging with alumni and the general public. A review of the unique appearance of a university on one or more lists demonstrates the diversity or lack thereof between the five rankings under consideration. Only Money Magazine, with its emphasis on perceived value, includes 115 institutions not evaluated elsewhere; while more than 53\% of the universities in our dataset appear on at least two of the indicated lists. This anchoring of universities among the ranking lists is consistent with previous research \cite{bowman2011anchoring} regarding adherence to the status quo (see Section 2.1). 

% \usepackage{booktabs}
\begin{table}[ht]
	\centering
	\caption{Rank Sequencing Using Spearman\textquotesingle s Footrule}
	\label{tab:RankSequencing}
	\resizebox{0.80\linewidth}{!}{%
		\begin{tabular}{@{}lrr@{}}
			\toprule
			University                                  & THE     & THE \\
			& Rank    & Ordered \\ 
			\midrule
			Stanford University                         & 3        & 1           \\
			Harvard University                          & 6        & 2           \\
			Princeton University                        & 7        & 3           \\
			Yale University                             & 12       & 4           \\
			University of California--Berkeley          & 13       & 5           \\
			Columbia University                         & 15       & 6           \\
			University of California--Los Angeles       & 16       & 7           \\
			University of Pennsylvania                  & 17       & 8           \\
			Cornell University                          & 18       & 9           \\
			Duke University                             & 20       & 10          \\
			University of San Diego						& ---	   & 112         \\
			Old Dominion University                     & ---      & 112         \\ 
			\bottomrule
		\end{tabular}%
	}
\end{table}

\subsection{Standardizing the Rank Positions}
Two of the ranking systems that contribute to our dataset bin universities alphabetically into groups after a certain threshold has been reached, resulting in tied ranking positions for those universities found lower on the list. After the first 200 individual rankings, THE places the remaining institutions ranked between 201 and 400 into bins of size 50 and then use bins of size 100 for ranks between 401 and 800. The ranking for each binned institution is the lowest number in the bin. All institutions listed alphabetically as ranked between 401 and 500 would be assigned rank 401. The rankings of ARWU are conducted similarly except ARWU starts to bin after the first 100 individual rankings.

One of the problems when comparing two ranked lists is that the items ranked in two particular lists are not identical, meaning items that appear in list A do not necessarily appear in list B. Fagin ~\cite{fagin2003comparing} introduced a new measure which extends Spearman\textquotesingle s Footrule by assigning a rank to the non-overlapping elements. For two rankings of size k, each element that appears in list A but does not appear in the list B (either totally missing from B or ranked at position [k]) is assigned rank k+1. For the purpose of our research, application of the footrule essentially places all  universities which are not ranked at the end of a respective list.  After removing the international entries, if any, the remaining institutions on each ranking list were sequentially ordered by rank as shown in Table ~\ref{tab:RankSequencing} using the THE rank as an example. The sequential ordering according to relative position was necessary because of differences in the number of U.S. institutions on each list (see Table ~\ref{tab:University Ranking Lists}), and the need to standardize ranking positions to obtain concordance between all lists.

\subsection{Computing Adjusted Reputation Rank}
One of our research goals is to compute an adjusted reputation rank. Therefore, we must avoid unduly penalizing an institution by including a low, raw ranking on a particular list in our ARR calculation; especially when the institution is referenced on just one or two of the named lists. To ensure that we incorporate different ranking perspectives in our evaluation, we average the ordered positional rankings from all ranking lists in our consolidated dataset to compute a mean reputation score which we then use to sequentially order the listed universities to obtain the adjusted reputation rank shown in Table \ref{tab:AdjustedReputationOrder}. Upon examination, we discovered that some schools which met the criteria to be ranked by Money Magazine based on \textit{value} performed differently using the criteria established by the other ranking systems. For example, Columbia University is consistently in the top-15 of the other four ranking systems while Money Magazine ranks the school considerably lower at position 52. As described later in Section 4.1, we computed rank-order correlation for each of the rankings. Table ~\ref{tab:RankvsRankKendall} shows that the rankings from Money Magazine are consistently weak to moderately correlated with all other ranking lists we consider. Therefore, we exclude the Money Magazine rankings from our computation of ARR. The 115 schools which appeared only on Money Magazine were placed in a non-ranked position at the end of ARWU, THE, and the lists from U.S. News. A standardized ranking position was then calculated using the methodology described in Section 3.2.

% \usepackage{booktabs}
\begin{table*}[ht]
	\centering
	\caption{Top-15 Universities Ranked by EEE}
	\label{tab:Top15ByEEE}
	\resizebox{0.7\linewidth}{!}{%
	\begin{tabular}{@{}lrrrr@{}}
		\toprule
		University & \begin{tabular}[c]{@{}r@{}}Undergraduate\\ Enrollment\end{tabular} &  \begin{tabular}[c]{@{}r@{}}Endowment,\\ Thousands \$\end{tabular} & \begin{tabular}[c]{@{}r@{}}Athletic\\ Expenditures, \$\end{tabular} & EEE \\ \midrule
		Ohio State University & 40,452 & 3,633,887 & 136,966,818 & 1 \\
		University of Texas & 36,072 & 3,341,835 & 152,853,239 & 2 \\
		Pennsylvania State University & 39,077 & 3,635,730 & 117,818,050 & 3 \\
		University of Michigan & 27,297 & 9,952,113 & 131,003,957 & 3 \\
		University of Wisconsin--Madison & 27,867 & 2,465,051 & 122,975,876 & 5 \\
		University of Florida & 29,577 & 1,550,000 & 130,772,416 & 6 \\
		Michigan State University & 35,038 & 2,274,813 & 89,491,630 & 7 \\
		University of Washington & 27,733 & 3,076,226 & 88,580,078 & 8 \\
		University of California--Los Angeles & 29,027 & 1,864,605 & 96,912,767 & 9 \\
		Indiana University & 31,161 & 1,974,215 & 81,161,423 & 10 \\
		University of California--Berkeley & 26,320 & 7,997,099 & 76,348,304 & 11 \\
		University of Illinois & 31,312 & 1,585,807 & 74,469,976 & 12 \\
		Purdue University & 28,382 & 2,397,902 & 66,164,834 & 13 \\
		University of Southern California & 17,898 & 4,709,511 & 105,919,366 & 14 \\
		University of Georgia & 25,259 & 1,004,987 & 101,559,307 & 15 \\ \bottomrule
	\end{tabular}%
}
\end{table*}

\subsection{Computing the Composite EEE Rank}
We identified several candidate attributes in order to determine which combination of quantifiable attributes might provide a good evaluation metric for our ranking system. We empirically selected a combination of web-based and other characteristics which might be calculated or retrieved from the Web: athletic expenditures, undergraduate enrollment, monetary value of the endowment, institution age, primary and secondary Twitter followers. We also combined several of these metrics into a composite ranking consisting of endowment, expenditures, and enrollment (EEE); metrics which are possible to collect from web-based sources. The top-15 universities as ranked by our EEE score are shown in Table ~\ref{tab:Top15ByEEE}.  Due to the broad range of values in the individual components, each of the enrollment, endowment and expenditures was normalized individually across the full dataset of 264 universities to obtain the same scale, from 0 to 1, then aggregated to obtain a sequential EEE ranking of the universities.

We chose to include the total expenditures for men\textquotesingle s and women\textquotesingle s sports as a measure of the institution\textquotesingle s commitment to branding and promoting the university as a whole. Further, we theorize whether the EEE score might serve as a viable proxy measure for a subset of our data, the NCAA Power Five, that we use later in Section 4.3 to assess the strength of UTE as a ranking attribute. The NCAA Power Five Conferences include the Southeastern Conference (SEC), Atlantic Coast Conference (ACC), Big Ten, Pac-12, and Big 12. The chosen conferences are composed of 65 flagship public and private universities who share excellent academic reputations, large endowments, and big budgets allocated for their athletic programs. These schools are representative of institutions that are playing at the highest level of NCAA competition and typically excel in two if not all three of the dimensions of enrollment, expenditures, and endowment.

% \usepackage{booktabs}
\begin{table*}[]
	\centering
	\caption{Union of the Top 15 Universities According to ARR and Top 15 According to UTE, sorted by ARR. UTE score is the sum of the primary and secondary followers.}
	\label{tab:AdjustedReputationOrder}
	\begin{tabular}{@{}lrrrrrrrr@{}}
		\toprule
		\multicolumn{1}{c}{University} & \begin{tabular}[c]{@{}r@{}}ARWU\\ Ordered\end{tabular} & \begin{tabular}[c]{@{}r@{}}THE\\ Ordered\end{tabular} & \begin{tabular}[c]{@{}r@{}}USNEWS\\ 2015\\ Ordered\end{tabular} & \begin{tabular}[c]{@{}r@{}}USNEWS\\ 2016\\ Ordered\end{tabular} & \begin{tabular}[c]{@{}r@{}}Mean\\ Reputation\\ Score\end{tabular} & \begin{tabular}[c]{@{}r@{}}Adjusted\\ Reputation\\ Rank\end{tabular} & \begin{tabular}[c]{@{}r@{}}UTE\\ Score\end{tabular} & \begin{tabular}[c]{@{}r@{}}UTE\\ Rank\end{tabular} \\ \midrule
		Harvard University & 1 & 2 & 1 & 1 & 1 & 1 & 4,562,501 & 1 \\
		Stanford University & 2 & 1 & 3 & 3 & 2 & 2 & 2,239,440 & 2 \\
		University of California--Berkeley & 3 & 5 & 2 & 2 & 3 & 3 & 474,901 & 19 \\
		Princeton University & 4 & 3 & 6 & 7 & 5 & 4 & 574,758 & 15 \\
		Columbia University & 5 & 6 & 5 & 5 & 5 & 4 & 759,574 & 7 \\
		University of California--Los Angeles & 7 & 7 & 4 & 4 & 6 & 6 & 394,815 & 28 \\
		Yale University & 6 & 4 & 9 & 8 & 7 & 7 & 808,461 & 4 \\
		University of Pennsylvania & 10 & 8 & 10 & 8 & 9 & 8 & 778,805 & 5 \\
		University of Washington & 9 & 13 & 7 & 6 & 9 & 8 & 274,674 & 44 \\
		University of Michigan & 11 & 11 & 7 & 10 & 10 & 10 & 671,277 & 12 \\
		Cornell University & 8 & 9 & 12 & 12 & 10 & 10 & 820,656 & 3 \\
		Duke University & 16 & 10 & 11 & 11 & 12 & 12 & 323,231 & 37 \\
		University of Minnesota & 15 & 23 & 16 & 17 & 18 & 16 & 631,046 & 13 \\
		Ohio State & 29 & 28 & 19 & 20 & 24 & 22 & 596,390 & 14 \\ 
		Pennsylvania State & 26 & 25 & 26 & 28 & 26 & 24 & 693,971 & 11 \\
		Arizona State & 36 & 112 & 45 & 45 & 60 & 59 & 770,711 & 6 \\
		\bottomrule
	\end{tabular}
\end{table*}

\subsection{Collecting University Demographic Data}
As a starting point for obtaining key institutional and demographic information for each university, we extracted (scraped) the associated website as listed on the university\textquotesingle s profile page maintained by the ranking list. We extracted information from multiple websites which included Division I conference membership from the National Collegiate Athletic Association (NCAA), athletic expenditures and endowment value from the National Center for Education Statistics\footnote{\url{https://nces.ed.gov/}}, profile data from Twitter, historical conference data from Sports Reference\footnote{\url{http://www.sports-reference.com/}}, primary and secondary Twitter account names from university homepages, undergraduate enrollment from the Integrated Postsecondary Education Data System\footnote{\url{https://nces.ed.gov/ipeds/}} (IPEDS) and founding dates from  DBpedia\footnote{http://wiki.dbpedia.org/}. For endowments that were attributed to a university system (e.g., University of Minnesota Foundation vs. University of Minnesota-Twin Cities), we used DBpedia to obtain the endowment value for the particular university present in the ranking lists to avoid overstating the endowment. Specific institutional data such as the founding date that could not be obtained from another already mentioned source was also resolved using web searches of DBpedia.

\begin{algorithm}
	\caption{Mining Official Twitter Accounts}
	\label{alg:Mining Twitter Accounts}
	\begin{algorithmic}[1]
		\State Let $h\gets homePageURI$ 
		\State Let $d\gets domain(h)$
		\State $primaryTwitterAccts\gets findOfficialTwitterAccounts(h,d)$
		\Function{findOfficialTwitterAccounts}{$H,$D}
		\State $foundAccountInd\gets false$
		\State $TwitterPrimary\gets nil$
		\State $W\gets ViewPageSource(H)$
		\Repeat
		
		\Comment{Search for anchor tag with href in the Twitter format}
		\State $A\gets anchorTag$		

		\State $user\gets TwitterRegexp(A)$
		\If{$user\equiv TwitterAccount$ }
		\State $profile\gets Twitter GET users(user)$
		\If{$domain(profileURI)\subset D}$
		
		\Comment{Twitter friends are the users an account follows}
		\State $TwitterPrimary\gets TwitterPrimary\cup profile$
		\State $friends\gets Twitter GET Friends(profile)$
		\State $TwitterPrimary\gets TwitterPrimary\cup friends$
		\State $foundAccountInd\gets true$
		\EndIf
		\EndIf
		\Until{$W\equiv nil}$
		
		\If{$foundAccountInd}$
		\State $UTE \gets 0$
		\For{i=1} {length(TwitterPrimary)}
		\State $primAcct\in TwitterPrimary(i)$
		\State $profile\gets Twitter GET Followers(primAcct)$
		\If{$domain(profileURI)\subset D}$
		\State $UTE\gets UTE+followers$
		\EndIf
		\EndFor
		\Else
		\State $searchResults\gets GoogleCustomSearch(h,"twitter")$
		\State $TwitterPrimary\gets searchResults(0)$
		\State $UTE \gets 0$
		\State $primAcct\gets TwitterPrimary(0)$
		\State $profile\gets Twitter GET Followers(primAcct)$
		\If{$domain(profileURI)\subset D}$
		\State $UTE\gets UTE+followers$
		\EndIf
		\EndIf
		\Return $UTE$
		
		\EndFunction
	\end{algorithmic}
\end{algorithm}

\subsection{Mining Official Twitter Accounts}
\begin{figure*}[ht]
	\includegraphics[width=0.95\linewidth]{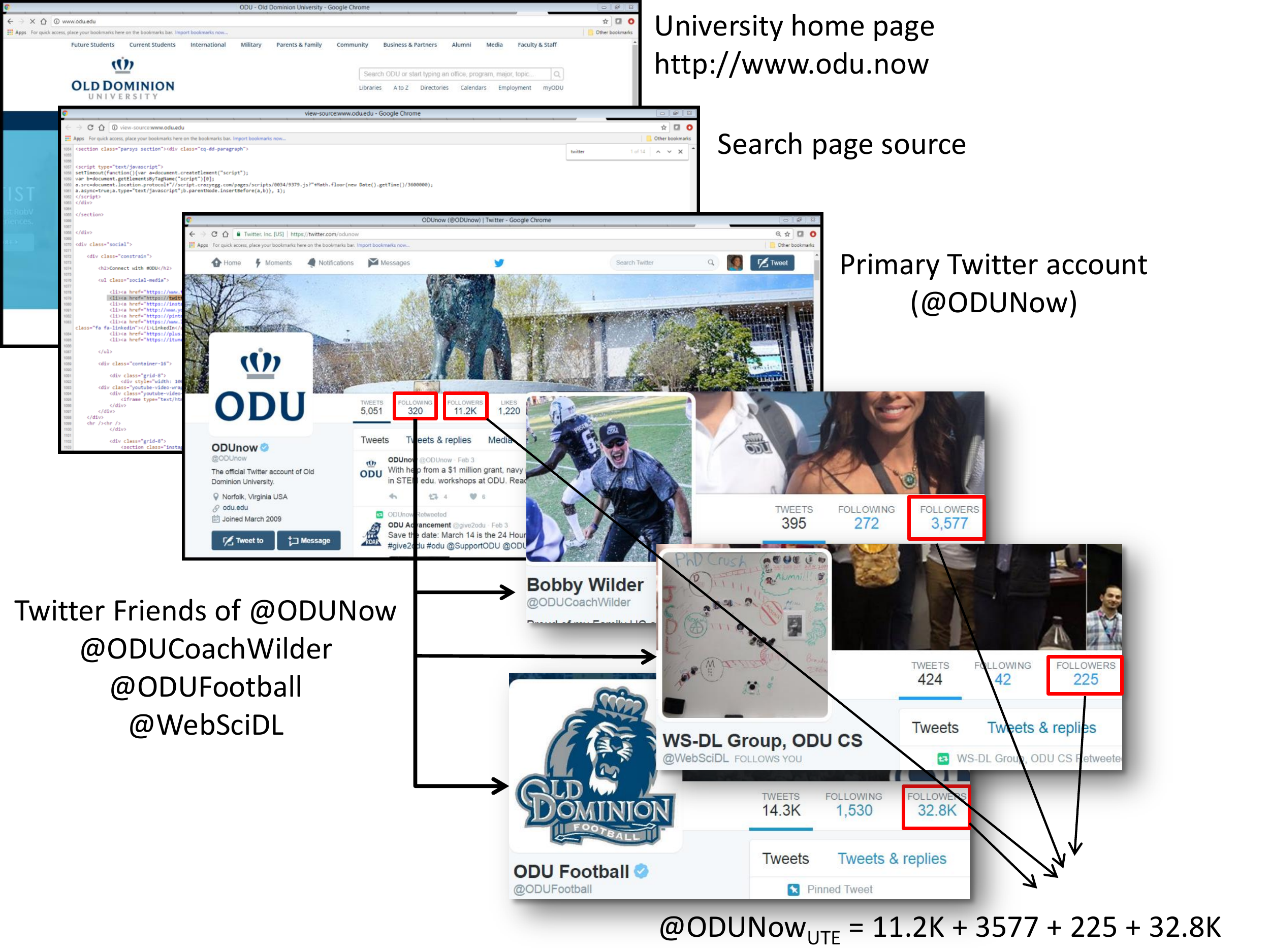}
	\caption{Mining Twitter Accounts.}
	\label{fig:Algorithm Visual Flow.}	
\end{figure*}

One of the proposed performance indicators for our dataset is constructed around a set of primary Twitter seed accounts for each university. For the present study, the presence of Twitter friends is also needed to bootstrap the discovery of affiliated, secondary Twitter accounts. The complete process for identifying these accounts and determining the value for UTE is shown in Algorithm \ref{alg:Mining Twitter Accounts} and described here. As illustrated in Figure ~\ref{fig:Algorithm Visual Flow.}, we start with the URI for the university\textquotesingle s homepage obtained from the detailed institutional profile information in the ranking lists. For each URI, we navigated to the associated webpage and searched the HTML source for links to valid Twitter handles. After examining the source anchor link text, we eliminated known false positives which were longer than 15 characters (Twitter limit for a valid screen name) or included /intent, /share, /tweet, /search or /hashtag in the URI which are directives to Twitter queries. Once the Twitter screen name was identified, the Twitter GET users/Show API was used to retrieve the URI from the profile of each user name. If the domain of the URI matched exactly or resolved to the known domain of the institution, we considered the account to be one of the university\textquotesingle s official, primary Twitter handles since the user had self-associated with the university via the URI reference. As an example, the user names $@$NBA, $@$DukeAnnualFund, $@$Duke\_MBB, and $@$DukeU were extracted from the page source of the Duke University homepage (www.duke.edu). However, only $@$DukeAnnualFund and $@$DukeU are considered official primary accounts because their respective URIs, annualfund.duke.edu and duke.edu, are in the same domain as the university.

As shown in Table  ~\ref{tab:Universities Without Twitter on Homepage.}, ten institutions did not have a Twitter account identified on the homepage as of August 2016, therefore, a primary official account could not be determined via our automated homepage search. For this subset only, we used the Google Custom Search Engine\footnote{\url{https://cse.google.com/cse/}} to initiate an X-ray search using the  keywords \textit{\textquotedblleft institution URI\textquotedblright ~AND \textquotedblleft  twitter\textquotedblright}. We accepted the top ranked result returned by Google, if any, as the official, primary Twitter account for the university. In the event that Google did not render a Twitter account in the search results, we manually searched for any remaining outstanding accounts using the search bar located on \url{http://twitter.com}. 

% \usepackage{booktabs}
\begin{table}[ht] 
	\centering
	\caption{Universities Without a Twitter Link on Their Homepage (as of August 2016)}
	\label{tab:Universities Without Twitter on Homepage.}
	\begin{tabular}{@{}ll@{}}
		\toprule
		\textbf{University}                     & \textbf{Twitter Screen Name} \\ \midrule
		University of Louisville                & @uofl                          \\
		University of South Carolina            & @uofsc                         \\
		University of Missouri                  & @mizzou                        \\
		University of North Carolina-Greensboro & @uncg                          \\
		Ball State                              & @ballstate                     \\
		University of Evansville                & @uevansville                   \\
		Fordham University                      & @fordhamnotes                  \\
		Marist College                          & @marist                        \\
		Portland State University               & @portland\_state               \\
		East Carolina                           & @eastcarolina                  \\ \bottomrule
	\end{tabular}
\end{table}

Colleges and universities have a reputation for being decentralized, with many departments operating independently of one another, maintaining a separate social media presence. However, we observed that only 24 of the 264 universities in our dataset promoted multiple, official Twitter accounts on their homepage. For the purpose of computing our UTE score, we want to consider the contribution of all university-affiliated Twitter accounts. Therefore, for each of the identified official, primary accounts, we obtained the full list of their Twitter friends, i.e., users that they follow. Again, we used the Twitter GET users/Show API to determine which of the friends could be included as secondary official Twitter accounts based on the URI in the profile (must have the same domain as the university). These secondary accounts might include the athletic teams, faculty members, and other university organizations. Once the primary and secondary accounts were identified, we used the Twitter GET followers/IDs API to retrieve and accumulate the follower count to form the UTE score for the university.

We launched our crawler to find all of the designated Twitter followers during the time period between June 15, 2016 and August 30, 2016. In total, we collected 1,087,000 user profiles. Approximately 9\% of all the user accounts we collected were protected at the profile owner\textquotesingle s request; allowing only their friends to view their profiles. Subsequently, we ignored these users in the computation of the UTE score because the underlying profile data is inaccessible using the Twitter API.  Once we calculated the UTE score, we then ranked each university, in sequential order, based on the score, as shown in Table ~\ref{tab:AdjustedReputationOrder}.

\section{Evaluation}
In this section we evaluate our UTE ranking by computing rank-order correlation with the adjusted reputation rank (Section 3.3) and the composite EEE rank (Section 3.4).  We also directly compare the rankings of individual universities for the full dataset and discuss the implications for universities in the NCAA Power Five conferences.
% \usepackage{booktabs}
\begin{table*}[]
	\centering
	\caption{Kendall\textquotesingle s Tau-b Correlation Between Ranking Lists and our Adjusted Reputation Rank (N=264)}
	\label{tab:RankvsRankKendall}
	\begin{tabular}{@{}rrrrrrrr@{}}
		\toprule
		& ARWU   & MONEY  & USNEWS2015 & USNEWS2016 & THE    & ARR      \\ 
		\midrule
		ARWU   			& 1      & 0.4191 & 0.8763 & 0.8565  & 0.7634 & 0.8533 \\
		MONEY  			& 0.4191 & 1      & 0.3761 & 0.3239  & 0.3504 & 0.3189 \\
		USNEWS2015      & 0.8763 & 0.3761 & 1      & 0.8787  & 0.7496 & 0.8542 \\
		USNEWS2016 		& 0.8565 & 0.3239 & 0.8787   & 1     & 0.7605 & 0.9375\\
		THE    			& 0.7634 & 0.3504 & 0.7496 & 0.7605  & 1      & 0.8285 \\
		ARR    			& 0.8533 & 0.3189 & 0.8542 & 0.9375  & 0.8285 & 1      \\
		\bottomrule
	\end{tabular}
\end{table*}

% \usepackage{booktabs}
\begin{table*}[h]
	\centering
	\caption{Kendall\textquotesingle s Tau-b Correlation Between Composite Rankings and UTE Rank for Institutions on Two or More Lists}
	\label{tab:RankvsRankKendall-UTE}
	\centering
	\begin{subtable}{0.5\columnwidth}	
		\begin{tabular}{@{}rrrr@{}}
			\toprule
			&EEE     & ARR    & UTE \\ 
			\midrule
			EEE  & 1      & 0.5310 & 0.5728 \\
			ARR  & 0.5310 & 1      & \textbf{0.6691} \\
			UTE  & 0.5728 & 0.6691 & 1 \\
			\bottomrule
		\end{tabular}
		\caption{Top 50 }
		\label{tab:RankvsRankKendall-UTE-Top50}
	\end{subtable}
	\centering
	\begin{subtable}{0.5\columnwidth}	
		\begin{tabular}{@{}rrrr@{}}
			\toprule
			&EEE     & ARR    & UTE \\ 
			\midrule
			EEE  & 1      & 0.5410 & 0.5620 \\
			ARR  & 0.5410 & 1      & \textbf{0.5920} \\
			UTE  & 0.5620 & 0.5920 & 1 \\
			\bottomrule
		\end{tabular}
		\caption{Top 100 }
		\label{tab:RankvsRankKendall-UTE-Top100}
	\end{subtable}
\centering	
\begin{subtable}{0.5\columnwidth}	
	\begin{tabular}{@{}rrrr@{}}
		\toprule
		&EEE     & ARR    & UTE \\ 
		\midrule
		EEE  & 1      & 0.5538 & 0.5960  \\
		ARR  & 0.5538 & 1      & \textbf{0.5967} \\
		UTE  & 0.5960 &  0.5967 & 1 \\
		\bottomrule
	\end{tabular}
	\caption{Top 141 }
	\label{tab:RankvsRankKendall-UTE-TwoOrMore}
\end{subtable}
	\centering	
	\begin{subtable}{0.5\columnwidth}	
		\begin{tabular}{@{}rrrr@{}}
			\toprule
			&EEE     & ARR    & UTE \\ 
			\midrule
			EEE  & 1      & 0.5969 & 0.6461 \\
			ARR  & 0.5969 & 1      & \textbf{0.6018} \\
			UTE  & 0.6461 & 0.6018 & 1 \\
			\bottomrule
		\end{tabular}
		\caption{All 264 }
		\label{tab:RankvsRankKendall-UTE-All264}
	\end{subtable}

\end{table*}

\subsection{Rank-Order Correlation}
 Since we know that the potential for tied rankings exists in our data, we used Kendall\textquotesingle s Tau-b ($\tau$) rank-order correlation to test for statistically significant (p $<$ 0.05), moderate (0.40 $<$ $\tau$ $\leq$ 0.60) or strong (0.60 $<$  $\tau$ $\leq$ 0.80) correlations between the individual ranking systems and our adjusted reputation rank. Table  ~\ref{tab:RankvsRankKendall} shows the respective inter-rank correlation measured in Kendall $\tau$. With $\tau$ values in the range of 0.3189 to 0.4191, the rankings on Money Magazine are weak to moderately correlated with all other ranking lists including our ARR. This range of $\tau$ values confirms our intuition that the disparate ranking criteria based on \textit{value} and the underlying goals of the Money Magazine system appropriately deem it an outlier among the other lists. We note a strong correlation, in the range of 0.7634 to 0.8787, between the remaining four lists which indicates that (1) the criteria traditionally used to rank universities based on academic excellence changes slowly thus resulting in minimal differentiation in the selected universities and (2) the relative ranking position of a particular university is anchored and does not vary significantly from year to year. The strong correlation of 0.8787 between subsequent lists found in the 2015 and 2016 rankings in U.S. News along with the addition of only three new entrants in 2016 (see Table ~\ref{tab:University Ranking Lists}) confirms this observation. The lack of variety between the U.S. News rankings is also consistent with the conclusions of Grewal et al. ~\cite{grewal2012university}, noted previously in Section 2.1, which indicated the high probability of a top-ranked university retaining its rank from year to year. Our adjusted reputation rank, with $\tau$ values in the range of 0.8285 to 0.9375, is strongly correlated with the rankings in ARWU, THE, and both years of USNEWS. Therefore, we conclude that ARR can be used as a representative proxy for any traditional ranking system. 

\begin{figure*}[]
	\begin{subfigure}[b]{0.30\textwidth}		
		\includegraphics[width=\textwidth]{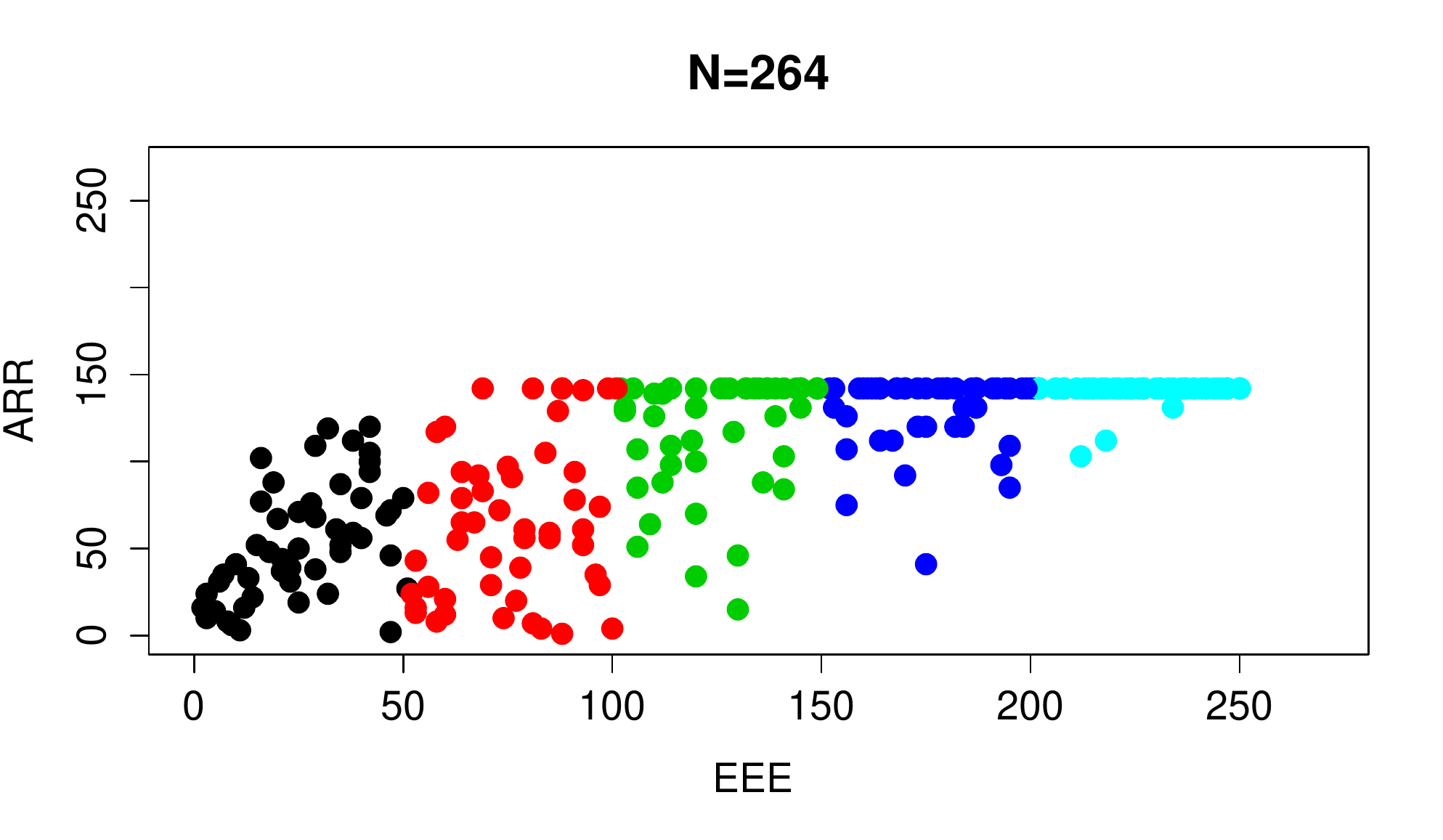}		
		\caption{ARR v. EEE}
		\label{fig:FullARRvEEE}
	\end{subfigure}
	\begin{subfigure}[b]{0.30\textwidth}
		\includegraphics[width=\textwidth]{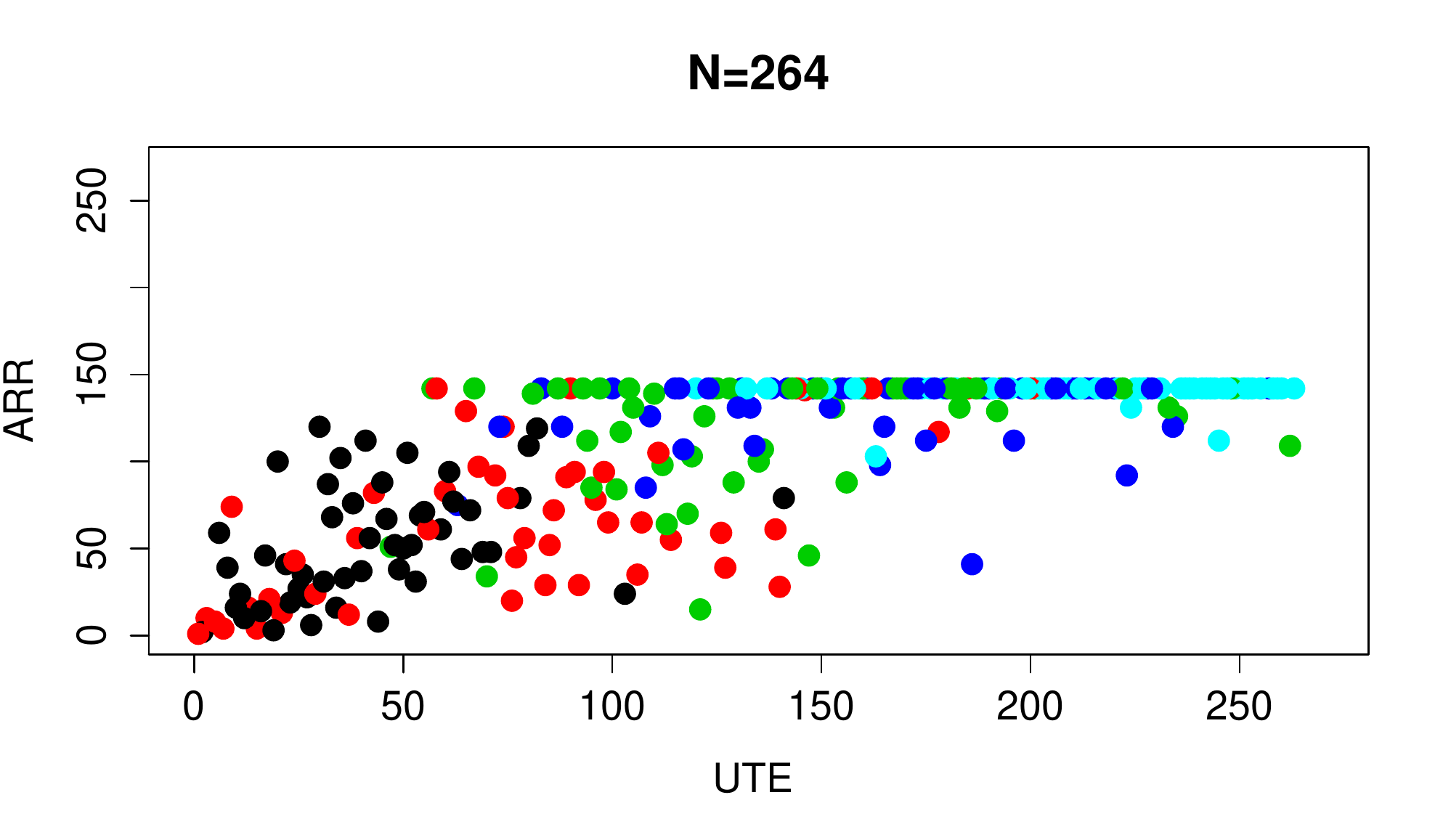}
		\caption{ARR v. UTE}
		\label{fig:FullARRvUTE}
	\end{subfigure}
	\begin{subfigure}[b]{0.30\textwidth}
		\includegraphics[width=\textwidth]{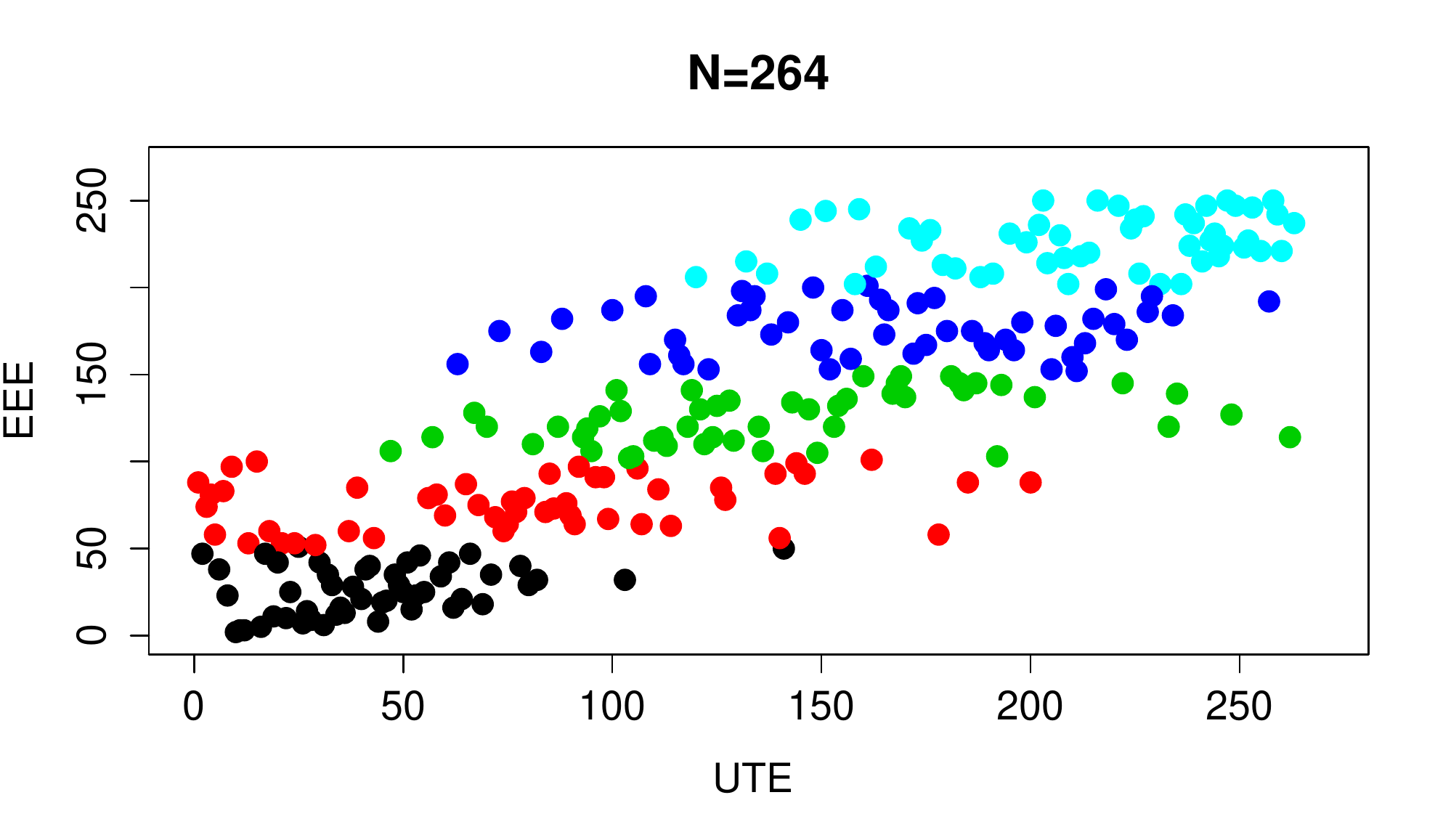}
		\caption{EEE v. UTE}
		\label{fig:FullEEEvUTE}
	\end{subfigure}
	\caption{Correlation of Composite Rankings (Full Dataset). Colors represent bins of the EEE rank from 1 to 264.}
	\label{fig:FullDatasetCorrelation}
\end{figure*}

\subsection{Composite Ranking Correlation with UTE}
In order to evaluate our UTE rank against the adjusted reputation rank and EEE rank, we again used Kendall\textquotesingle s Tau-b ($\tau$) rank-order correlation to test for statistically significant (p $<$ 0.05), moderate (0.40 $<$ $\tau$ $\leq$ 0.60) or strong (0.60 $<$  $\tau$ $\leq$ 0.80) correlations. Using ARR as the ranking criteria, we selected the top-50, top-100, top-141 ranked on two or more lists, and all 264 universities in our dataset. As shown in Table ~\ref{tab:RankvsRankKendall-UTE-Top50}, we found with a $\tau$ value of 0.6691, UTE is most strongly correlated with the ARR for the top-50 institutions followed closely by EEE at 0.5728. We must note the majority of the universities in the top-50 of any ranking list are usually members of the Ivy League or large schools with highly recognizable athletic programs like those in the Power Five (e.g., Ohio State, Penn State) so we might expect similarities in the metrics that comprise EEE. The correlation between UTE and ARR decreases slightly for the top-100, but persists to indicate a strong correlation, $\tau$ = 0.6018, when we examine the full dataset in Table ~\ref{tab:RankvsRankKendall-UTE-All264}. Our goal is to maximize the use of web-based metrics, therefore, choosing UTE over EEE should provide similar ranking results regardless of the size of the list. We conclude that primary and secondary Twitter followers, as we have defined for UTE, presents a strong metric for ranking and assessing the reputation of a university. 

To further investigate the correlation of ARR, UTE, and EEE, we show scatterplots in Figure ~\ref{fig:FullDatasetCorrelation} of the combinations of the three rankings for all 264 universities. The colors represent bins of the EEE rank, which can be directly seen in Figure ~\ref{fig:FullARRvEEE}.  As discussed in Section 3.3, the 115 schools that appeared exclusively  on the Money Magazine list were binned and all assigned a rank of 142 on the ARR. Note that all of the universities in the first bin of EEE (black dots) are ranked below 150 in ARR, suggesting that universities with high enrollments, endowments, and/or athletic budgets also have high academic rank. Figure ~\ref{fig:FullARRvUTE} (ARR vs. UTE) shows that there are several universities that have larger Twitter followings than can be explained just by academic rank (i.e., UTE rank is higher than ARR rank). Most of these rankings fall into the first bin of EEE, which could explain the increase in Twitter following. Twitter engagement provides an inexpensive means for smaller schools to reach a large audience, potentially enhancing their reputations. Figure ~\ref{fig:FullARRvUTE} also shows that there are several smaller schools (in the last EEE bin, cyan dots) that have larger Twitter followings than their academic rank (not ranked in ARR) or EEE would explain. These schools may be making a concerted effort to enhance their profile and could potentially move into the standard academic rankings in the future. This would be an interesting avenue for future study. Finally, Figure ~\ref{fig:FullEEEvUTE} shows EEE vs. UTE, which indicates that as expected, universities with more financial resources tend to have larger Twitter followings, though there are still some universities in the lower EEE bins that have significant Twitter followings.

\begin{figure*}[]
	\begin{subfigure}[b]{0.30\textwidth}		
		\includegraphics[width=\textwidth]{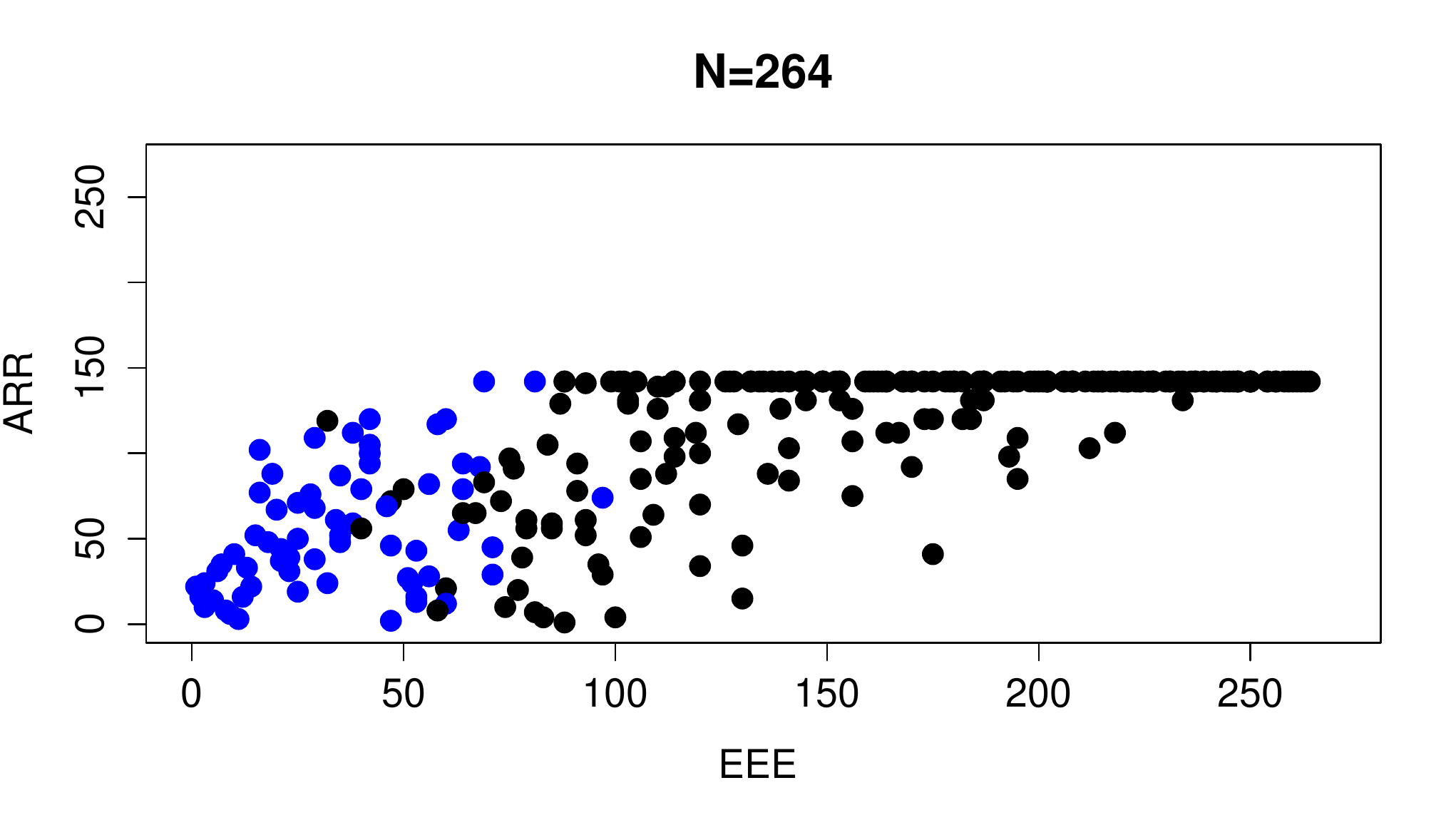}		
		\caption{ARR v. EEE}
		\label{fig:FullARRvEEE-Color}
	\end{subfigure}
	\begin{subfigure}[b]{0.30\textwidth}
		\includegraphics[width=\textwidth]{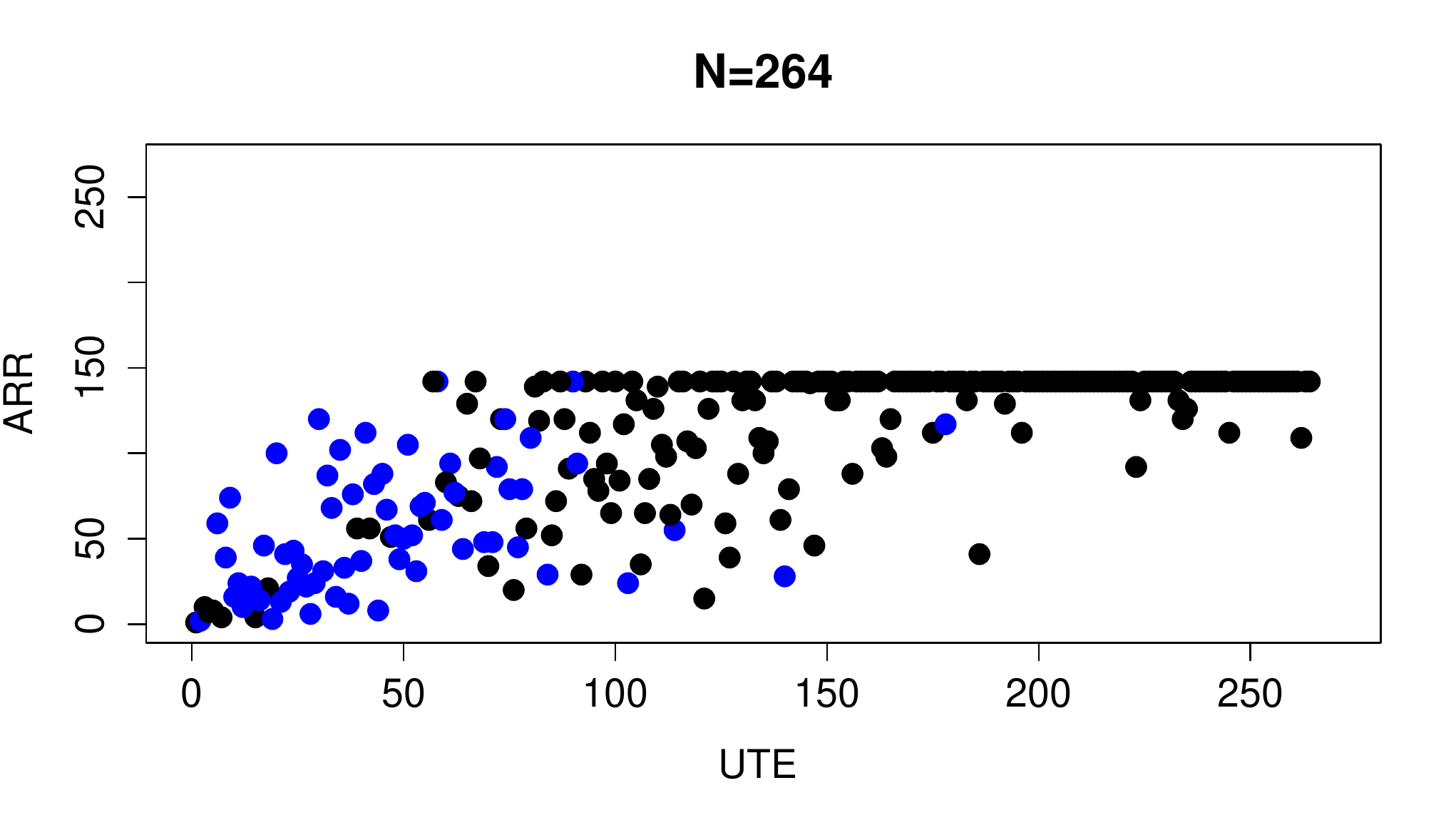}
		\caption{ARR v. UTE}
		\label{fig:FullARRvUTE-Color}
	\end{subfigure}
	\begin{subfigure}[b]{0.30\textwidth}
		\includegraphics[width=\textwidth]{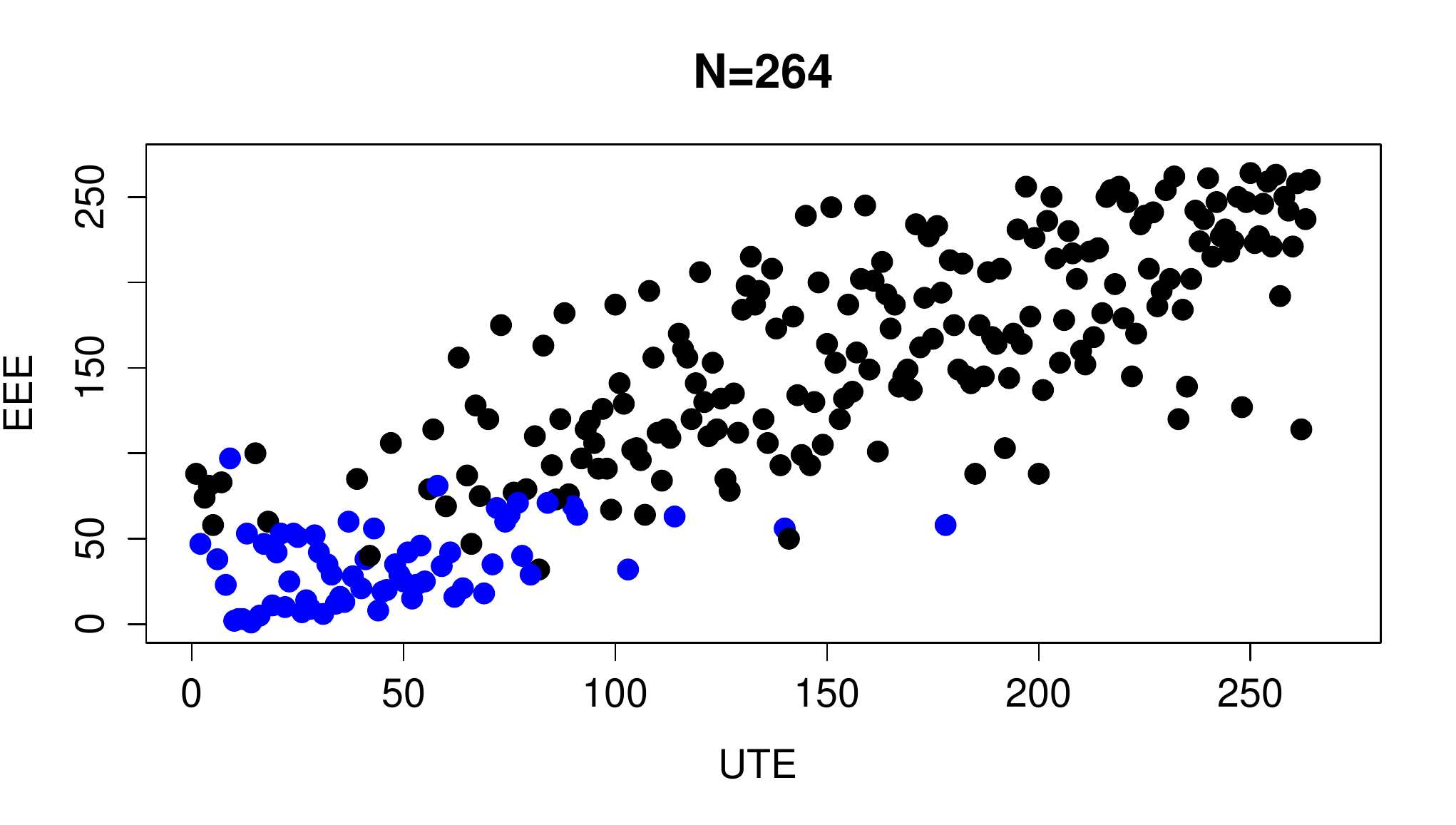}
		\caption{EEE v. UTE}
		\label{fig:FullEEEvUTE-Color}
	\end{subfigure}
	\caption{Correlation of Composite Rankings (Full Dataset). Blue dots represent member institutions in the Power Five.}
	\label{fig:FullDatasetCorrelationColorPower5}
\end{figure*}

\subsection{Correlation Between the NCAA Power Five}
We use the fraternity of the schools in the Power Five to more closely examine the collective ranking correlation of these conferences based on their 2016 membership. Within the complete data set, we observed that 55 out of the 65 Power Five member institutions (84.6\%) were ranked within the top-100 positions based on the ARR rank. Further, we found that all 65 schools (100\%) were ranked within the top-100 positions based on the EEE rank. The latter observation is consistent with the strong correlation between EEE and UTE, $\tau$ = 0.6461, that we determined in Table ~\ref{tab:RankvsRankKendall-UTE-All264} and is consistent with our intuition that large schools with ample financial resources would attract more Twitter followers. Figure ~\ref{fig:FullDatasetCorrelationColorPower5} highlights the relationships between the Power Five and the various metrics by repeating the same charts from Figure ~\ref{fig:FullDatasetCorrelation} but with members of the Power Five shown in blue. 

We noted several similarities which were indicative of the ten schools (15.4\%) that were ranked outside of the top tier for ARR. Notably both Texas Christian and Mississippi State are the only schools which were not ranked on two or more of the ranking lists. Both schools also fall significantly below the mean values for the Power Five in terms of undergraduate enrollment ($\approx$ 21,000), endowment value ($\approx$ \$2.3B), and athletic expenditures ($\approx$ \$90M), placing them at the bottom of the EEE ranking. On the other hand, Wake Forest is the smallest institution in the Power Five, but the school garners an academic reputation (ARR=45) that cannot be readily explained by its comparatively low EEE ranking (EEE=97). 

We also note four schools that fall within the bottom 50\% of UTE. In particular, the University of Louisville could achieve a considerable boost in UTE ranking ($\approx$ 107,000 followers) if the Twitter account used by the athletic department (@GoCards) would reference the primary URI of the university rather than its own domain (~\url{http://gocards.com}). We discovered 284 primary and secondary accounts followed by Georgia Tech, however only four of these could be considered official, because 150 of 280 secondary accounts did not include a URI in the profile bio. A similar scenario was noted for Oregon State where 271 of the 341 secondary accounts did not include a URI. While we identified 74 official accounts for the University of Pittsburgh, as was the case with Louisville, $\approx$ 140,000 underreported secondary Twitter accounts are associated with university sports. We discovered the Twitter followers of Wake Forest are bolstered significantly by a single celebrity professor, Melissa Harris Perry, who in addition to her faculty position previously hosted a weekly news style program on US television. More than 80\% of the Wake Forest UTE score is attributed to the verified $@$MHarrisPerry Twitter account which has more than 600,000 followers. 

In Appendix A, we note the diverse, though not exhaustive, spectrum of unique university domains found among secondary Twitter accounts of the NCAA Power Five. Upon visual inspection of the web content of each domain, we find they are related to the university in some capacity (e.g., sports, clubs), but do not conform to our domain association rule. The omission of the UTE for the associated secondary Twitter accounts can, in some cases, significantly lower our calculation of UTE score. For those under performing universities, in terms of Twitter followers, inclusion of more domains would elevate the UTE rank of the university and likely present a stronger correlation of Kendall\textquotesingle s Tau-b ($\tau$) than was noted in Table ~\ref{tab:RankvsRankKendall}. We did not attempt to identify additional secondary domains for the entire set of 264 universities in our dataset. This exercise would be manually intensive and counter to our stated goal of automated data collection.

\section{Discussion}
As noted during our own collection efforts, the quality and availability of the data chosen as performance indicators can impede the efficiency of constructing of a gold standard data set. Manual correction can improve the data collection, but is expensive and is not conducive to reproducible research. We observed that institutions themselves do not maintain a complete listing of all official Twitter accounts as noted by the number of undiscovered and undocumented accounts we extracted during a secondary search. We must also acknowledge the impact of celebrity professors and verified accounts (e.g., Melissa Harris Perry). Given the small number of verified accounts among our official Twitter profiles, we contend that celebrity faculty members might be equated to the influence of Nobel Prize laureates; an indicator which is used by some ranking systems. We did not address known issues with bots and spam accounts which may over inflate the stated number of Twitter followers which is the primary component of our UTE score (e.g., ~\cite{BotOrNot:2016}) . We also understand that our methodology constrains universities to a single official hostname which can deflate the UTE score as Twitter accounts that reference other university-owned domains are omitted. Based on our research assumptions, we observed that enrollment does not necessarily increase the Twitter followers needed to compute UTE. Universities are not taking the opportunity to advertise their Twitter accounts and are at times promoting other entities on their homepage. This observation necessitated the need to expand the follower network as we have defined. Schools with highly visible sports programs, like those in the Power Five, tend to have more Twitter followers as the public is more aware of the university\textquotesingle s overall brand. In general, the perceived reputation of any university is impacted less by metrics which are intrinsic to the institution, but intangibles that translate into more impressions or brand awareness by the public and constituents. This parallels the assertions in prior research ~\cite{leavitt2009influentials, preussler2010managing} which contends that popular entities are more likely to attract more followers (see Section 2.3).

\section{Conclusions and Future Work}
We examined and ranked a set of 264 U.S. universities extracted from the NCAA Division I  membership and lists published in U.S. News, Times Higher Education, Academic Ranking of World Universities and Money Magazine using an adjusted reputation rank which we compared to our University Twitter Engagement score; the friend and extended follower network of primary and affiliated secondary Twitter accounts referenced on a university\textquotesingle s home page. When compared to our adjusted reputation rank for all 264 represented universities, we noted a strong correlation, $\tau$=0.6018, with UTE. We conclude that our UTE rank is comparable to those presented in other academic-based ranking systems, however, we present a low-cost data acquisition methodology using only web-based artifacts. UTE also offers a distinct advantage because (1) it can be calculated on-demand and (2) it promotes diversity in the ranking lists as any university with a Twitter account can be given a UTE rank. These results are highly reproducible as they are derived from social media and obtained using a publicly accessible Twitter API.  A similar aggregation strategy might also be applied to other popular social platforms such as Instagram or YouTube. The use of a web-based API allows our results to be calculated on a near-real time basis rather than annually which is the norm for other ranking systems. 

The use of web metrics might also provide an incentive for institutions to increase their web presence as way to further engage with constituents and the general public. Social media allows us to measure another proxy for reputation; how the universities and the public engage with one another. The universities themselves have to decide whether this kind of outreach is important and invest in it, and the public needs to be interested enough to follow them.

Our study is subject to a number of limitations that present opportunities for future work. Campbell's and Goodhart\textquotesingle s law suggest that if UTE becomes popular, institutions may seek to artificially increase their Twitter followers in order to increase their ranking.  Future work could include only the Twitter accounts of real people. In order to obtain a more complete set of official Twitter accounts, the domain associated with the account URI could be expanded to include all registered domains for the university. Additional research might also broaden the scope of our study to include both U.S. and international universities. It might also be advantageous to subject the observations made in this paper to a temporal analysis to ascertain whether the UTE rankings, at least for those in the upper echelon, persist over time and to look for non-linear spikes in Twitter followers which may indicate artificial manipulation. 

\bibliography{msBiblio}
%\end{document}  % This is where a 'short' article might terminate

\bibliographystyle{ACM-Reference-Format}

%Appendix of Supporting Tables
\clearpage
%https://tex.stackexchange.com/questions/127230/table-caption-spanning-two-columns
%https://tex.stackexchange.com/questions/301227/how-to-make-table-split-in-two-pages
\onecolumn
\appendix
\section{Supporting Tables}
\begin{center}
\begin{longtable}{@{}llrlrr@{}}
	\caption{Underreported UTE for NCAA Power Five Where the URL Does Not Conform to Domain Rules}
	\label{tab:SecondaryDomains-1}\\
	\toprule
		University & \begin{tabular}[c]{@{}l@{}}Homepage\\ Domain\end{tabular} & \begin{tabular}[c]{@{}r@{}}Unique\\ Secondary\\ Domains\end{tabular} & \begin{tabular}[c]{@{}l@{}}Sampling of\\ Secondary\\ Domains\end{tabular} & \begin{tabular}[c]{@{}r@{}}Associated\\ Secondary\\ Twitter\\ Accounts\end{tabular} & \begin{tabular}[c]{@{}r@{}}Associated\\ Secondary\\ UTE\end{tabular}  \\ \midrule
		\endfirsthead
		{\tablename\ \thetable\ -- \textit{Continued from previous page}} \\
	\toprule
		University & \begin{tabular}[c]{@{}l@{}}Homepage\\ Domain\end{tabular} & \begin{tabular}[c]{@{}r@{}}Unique\\ Secondary\\ Domains\end{tabular} & \begin{tabular}[c]{@{}l@{}}Sampling of\\ Secondary\\ Domains\end{tabular} & \begin{tabular}[c]{@{}r@{}}Associated\\ Secondary\\ Twitter\\ Accounts\end{tabular} & \begin{tabular}[c]{@{}r@{}}Associated\\ Secondary\\ UTE\end{tabular}  \\ \midrule
		\endhead		
\\ \hline \multicolumn{6}{r}{\textit{Continued on next page}} \\
\endfoot
\\ \hline
\endlastfoot		
	
		Arizona State University & asu.edu & 92 & \begin{tabular}[c]{@{}l@{}}thesundevils.com\\ asufoundation.org\end{tabular} & 138 & 498,097 \\
		Auburn University & auburn.edu & 15 & \begin{tabular}[c]{@{}l@{}}auburntigers.com\\ auburnalabama.org\\ auburn.collegiatelink.net\end{tabular} & 43 & 809,923 \\
		Baylor University & baylor.edu & 25 & \begin{tabular}[c]{@{}l@{}}baylorbears.com\\ baylormenschoir.org\end{tabular} & 62 & 308,520 \\
		Boston College & bc.edu & 68 & \begin{tabular}[c]{@{}l@{}}bceagles.com\\ eaglemunc.org\end{tabular} & 68 & 175,227 \\
		Clemson University & clemson.edu & 8 & \begin{tabular}[c]{@{}l@{}}clemsontigers.com\\ clemsongreeklife.com\end{tabular} & 14 & 224,964 \\
		Duke University & duke.edu & 48 & \begin{tabular}[c]{@{}l@{}}goduke.com\\ dukeoutoftheblue.org\end{tabular} & 78 & 1,046,188 \\
		Florida State University & fsu.edu & 48 & \begin{tabular}[c]{@{}l@{}}seminoles.com\\ fsulacrosse.com\end{tabular} & 79 & 490,251 \\
		Georgia Tech & gatech.edu & 1 & ramblinwreck.com & 1 & 2,305 \\
		Indiana University & iub.edu & 18 & \begin{tabular}[c]{@{}l@{}}iuhoosiers.com\\ hoosierhockey.com\end{tabular} & 52 & 164,536 \\
		Iowa State University & iastate.edu & 16 & \begin{tabular}[c]{@{}l@{}}cyclones.com\\ iastate.kappadelta.org\end{tabular} & 32 & 261,369 \\
		Kansas State University & k-state.edu & 4 & \begin{tabular}[c]{@{}l@{}}stateproud.org\\ wildcatsforever.com\end{tabular} & 4 & 22,492 \\
		Louisiana State & lsu.edu & 70 & \begin{tabular}[c]{@{}l@{}}lsusports.net\\ deltagammalsu.com\\ tigertv.tv\end{tabular} & 124 & 1,205,973 \\
		Michigan State University & msu.edu & 27 & \begin{tabular}[c]{@{}l@{}}msuspartans.com\\ spartanband.net\\ msupress.org\end{tabular} & 40 & 582,386 \\
		Mississippi State University & msstate.edu & 10 & \begin{tabular}[c]{@{}l@{}}msufoundation.com\\ msubulldogbash,com\end{tabular} & 13 & 27,072 \\
		North Carolina State University & ncsu.edu & 10 & \begin{tabular}[c]{@{}l@{}}wolfpackclub.comncsu\\ panhellenic.com\end{tabular} & 10 & 32,847 \\
		Northwestern University & northwestern.edu & 36 & \begin{tabular}[c]{@{}l@{}}northwestern.zetataualpha.org\\ northwestern.freshu.io\\ wildcatexpressdelivery.com\end{tabular} & 36 & 102,897 \\
		Ohio State University & osu.edu & 28 & \begin{tabular}[c]{@{}l@{}}osurugby.com\\ ohiostatebuckeyes.com\end{tabular} & 65 & 718,025 \\
		Oklahoma State University & osu.okstate.edu & 35 & \begin{tabular}[c]{@{}l@{}}ucatokstate.org\\ okstatecru.com\\ cowboywrestlilngclub.org\end{tabular} & 44 & 51,474 \\
		Oregon State University & oregonstate.edu & 9 & \begin{tabular}[c]{@{}l@{}}osubeavers.com\\ beaverblitz.com\end{tabular} & 18 & 150,641 \\
		Pennsylvania State University & psu.edu & 71 & \begin{tabular}[c]{@{}l@{}}gopsusports.com\\ ladylions.com\\ pennstategleeclub.com\end{tabular} & 121 & 767,708 \\		
		Purdue University & purdue.edu & 181 & \begin{tabular}[c]{@{}l@{}}purduesports.com\\ purduealumni.org\end{tabular} & 296 & 426,586 \\
		Rutgers University & rutgers.edu & 15 & \begin{tabular}[c]{@{}l@{}}scarletknights.com\\ rutgersalumni.org\end{tabular} & 19 & 80,939 \\
		Stanford University & stanford.edu & 20 & \begin{tabular}[c]{@{}l@{}}gostanford.com\\ cardinalredfootball.com\end{tabular} & 44 & 320,845 \\
		Syracuse University & syr.edu & 39 & \begin{tabular}[c]{@{}l@{}}syracuse.com\\ dailyorange.com\end{tabular} & 58 & 402,634 \\
		Texas A\&M University & tamu.edu & 49 & \begin{tabular}[c]{@{}l@{}}aggieathletics.com\\ tamuweather.org\end{tabular} & 64 & 331,272 \\
		Texas Christian University & tcu.edu & 38 & \begin{tabular}[c]{@{}l@{}}tcuftw.com\\ tcugammaphibeta.com\end{tabular} & 56 & 218,283 \\
		Texas Tech University & ttu.edu & 66 & \begin{tabular}[c]{@{}l@{}}redraidersports.org\\ ttulibrary.weebly.com\end{tabular} & 108 & 98,267 \\
		University of Alabama & ua.edu & 14 & \begin{tabular}[c]{@{}l@{}}uaemba.com\\ crimsontidefoundation.org\end{tabular} & 19 & 101,359 \\
		University of Arizona & arizona.edu & 19 & \begin{tabular}[c]{@{}l@{}}arizonawildcats.com\\ arizonawrugby.com\end{tabular} & 50 & 690,450 \\
		University of Arkansas & uark.edu & 14 & \begin{tabular}[c]{@{}l@{}}arkansasrazorbacks.com\\ uark.swe.org\end{tabular} & 36 & 886,112 \\
		University of California--Berkeley & berkeley.edu & 42 & \begin{tabular}[c]{@{}l@{}}berkeleystudentfoodcollective.org\\ ucberkeleymcc.tumblr.com\end{tabular} & 48 & 92,604 \\
		University of California--Los Angeles & ucla.edu & 22 & \begin{tabular}[c]{@{}l@{}}uclabruins.com\\ uclaextension.com\end{tabular} & 52 & 290,477 \\
		University of Colorado & colorado.edu & 6 & \begin{tabular}[c]{@{}l@{}}coloradocollege.edu\\ buffalobicycleclassic.com\end{tabular} & 6 & 16,002 \\
		University of Florida & ufl.edu & 11 & \begin{tabular}[c]{@{}l@{}}floridagators.com\\ gatorszone.com\\ ufl.collegiatelink.net\end{tabular} & 37 & 784,674 \\
		University of Georgia & uga.edu & 24 & \begin{tabular}[c]{@{}l@{}}ugafootballlive.com\\ ugabookstore.com\end{tabular} & 31 & 427,758 \\
		University of Illinois & illinois.edu & 66 & \begin{tabular}[c]{@{}l@{}}fightingillini.com\\ illinihockey.com\end{tabular} & 140 & 452,158 \\
		University of Iowa & uiowa.edu & 22 & \begin{tabular}[c]{@{}l@{}}hawkeyesports.com\\ sigmanuiowa.org\end{tabular} & 47 & 394,801 \\
		University of Kansas & ku.edu & 78 & \begin{tabular}[c]{@{}l@{}}kuathletics.com\\ jayhawkhockey.com\end{tabular} & 128 & 592,790 \\
		University of Kentucky & uky.edu & 13 & \begin{tabular}[c]{@{}l@{}}wildcatworld.com\\ uky.kappa.org\end{tabular} & 16 & 30,676 \\
		University of Louisville & louisville.edu & 8 & \begin{tabular}[c]{@{}l@{}}louisville.n.rivals.com\\ uclublouisville.org\end{tabular} & 11 & 89,035 \\
		University of Maryland & maryland.edu & 5 & \begin{tabular}[c]{@{}l@{}}terrapinstationmd.com\\ umaryland.edu\end{tabular} & 7 & 17,523 \\
		University of Miami & miami.edu & 6 & \begin{tabular}[c]{@{}l@{}}hurricanesports.com\\ themiamihurricane.com\end{tabular} & 24 & 251,879 \\
		University of Michigan & umich.edu & 5 & \begin{tabular}[c]{@{}l@{}}umich.com\\ wolverinesforlife.org\end{tabular} & 5 & 13,065 \\
		University of Minnesota & twin-cities.umn.edu & 63 & \begin{tabular}[c]{@{}l@{}}gophersports.com \\ minnesotaalumni.org\end{tabular} & 87 & 241,487 \\
		University of Mississippi & olemiss.edu & 58 & \begin{tabular}[c]{@{}l@{}}olemisssports.com\\ omrebelnation.com\end{tabular} & 119 & 823,676 \\
		University of Missouri & missouri.edu & 2 & mutigers.com & 11 & 23,629,273 \\
		University of Nebraska & unl.edu & 2 & \begin{tabular}[c]{@{}l@{}}unl.kappadelta.org \\ unlbookstore.com\end{tabular} & 35 & 28,970 \\
		University of North Carolina & unc.edu & 79 & \begin{tabular}[c]{@{}l@{}}dailytarheel.com\\ uncc.edu\end{tabular} & 108 & 331,103 \\
		University of Notre Dame & nd.edu & 11 & \begin{tabular}[c]{@{}l@{}}und.com\\ notredamefootballschedule.net\end{tabular} & 15 & 55,875 \\
		University of Oklahoma & ou.edu & 9 & \begin{tabular}[c]{@{}l@{}}soonersports.com\\ soonersports.tv\end{tabular} & 35 & 977,324 \\
		University of Oregon & uoregon.edu & 25 & \begin{tabular}[c]{@{}l@{}}goducks.com\\ uoduckstore.com\end{tabular} & 62 & 467,767 \\
		University of Pittsburgh & pitt.edu & 4 & \begin{tabular}[c]{@{}l@{}}pittsburghpanthers.com\\ pittpanthersgameday.com\end{tabular} & 21 & 140,584 \\
		University of South Carolina & sc.edu & 24 & gamecocksonline.com libraries.usc.edu & 72 & 1,003,376 \\
		University of Southern California & usc.edu & 110 & \begin{tabular}[c]{@{}l@{}}usctrojans.com \\ uscimpact.org\end{tabular} & 140 & 400,268 \\
		University of Tennessee & utk.edu & 36 & \begin{tabular}[c]{@{}l@{}}volnation.com\\ volsconnect.com\end{tabular} & 34 & 477,843 \\
		University of Texas & utexas.edu & 19 & \begin{tabular}[c]{@{}l@{}}longhornnetwork.com \\ utexasbands.org\end{tabular} & 22 & 196,330 \\
		University of Utah & utah.edu & 22 & \begin{tabular}[c]{@{}l@{}}utahutes.com\\ utefans.net\end{tabular} & 42 & 281,367 \\
		University of Virginia & virginia.edu & 8 & \begin{tabular}[c]{@{}l@{}}virginiasports.com\\ cavalierdaily.com\end{tabular} & 32 & 316,827 \\
		University of Washington & washington.edu & 7 & \begin{tabular}[c]{@{}l@{}}gohuskies.com \\ graduatewashington.org\end{tabular} & 43 & 252,407 \\
		University of Wisconsin & wisc.edu & 72 & \begin{tabular}[c]{@{}l@{}}uwbadgers.com\\ badgernation.com\end{tabular} & 118 & 978.770 \\
		Vanderbilt University & vanderbilt.edu & 20 & \begin{tabular}[c]{@{}l@{}}vucommodores.com\\ vanderbilt.org\end{tabular} & 30 & 252,776 \\
		Virginia Tech & vt.edu & 36 & \begin{tabular}[c]{@{}l@{}}hokiesports.com\\ hokienation.us\end{tabular} & 54 & 274,400 \\
		Wake Forest & wfu.edu & 22 & \begin{tabular}[c]{@{}l@{}}deaconclub.com \\ demondivaswfu.com\end{tabular} & 32 & 13,098 \\
		Washington State University & wsu.edu & 54 & \begin{tabular}[c]{@{}l@{}}wsucougars.com \\ wsujobs.com\end{tabular} & 111 & 210,679 \\
		West Virginia University & wvu.edu & 26 & \begin{tabular}[c]{@{}l@{}}wvusports.com\\ wvufootball.com\end{tabular} & 38 & 434,385 \\ 
\end{longtable}
\end{center}

\end{document}